\documentclass{article}
\usepackage{spconf,amsmath,graphicx}
\usepackage[hidelinks]{hyperref}
\usepackage[T1]{fontenc}
\usepackage{multirow}
\usepackage{booktabs} 
\usepackage{eso-pic}

\title{Estimating Respiratory Effort From Nocturnal Breathing Sounds For Obstructive Sleep Apnoea Screening}
\name{Xiaolei Xu$^{1}$, Chaoyue Niu$^{1}$, Guy J. Brown$^1$, Hector Romero$^2$ and Ning Ma$^1$
\thanks{The authors would like to thank PFL Healthcare (UK) for providing the sleep recordings. Xiaolei Xu is partially supported by the Ph.D. studentship from Passion For Life Healthcare (UK) Ltd. The study was funded by UKRI MRC impact acceleration accounts (grant 182731), NIHR Great Ormond Street Hospital Biomedical Research Centre (grant 187217), and Innovate UK Open Grant 26767.}
}
\address{$^1$School of Computer Science, University of Sheffield, Sheffield, S1 4DP, UK.\\
$^2$Passion for Life Healthcare, Chester, CH1 2NP, UK\\
Email: \{xxu97, c.niu, g.j.brown, n.ma\}@sheffield.ac.uk, hector.romero@passionforlife.com}

\begin{document}
\ninept

\AddToShipoutPictureFG*{%
  \AtPageLowerLeft{%
    \raisebox{1.5cm}{%
      \makebox[\paperwidth][c]{%
        \parbox{0.92\paperwidth}{\centering\footnotesize
        \copyright~2025 IEEE. Personal use of this material is permitted. Permission from IEEE must be obtained for all other uses,\\
        in any current or future media, including reprinting/republishing this material for advertising or promotional purposes,\\
        creating new collective works, for resale or redistribution to servers or lists, or reuse of any copyrighted component of this work in other works.}%
      }%
    }%
  }%
}
\maketitle
\begin{abstract}
Obstructive sleep apnoea (OSA) is a prevalent condition with significant health consequences, yet many patients remain undiagnosed due to the complexity and cost of overnight polysomnography. Acoustic-based screening provides a scalable alternative, yet performance is limited by environmental noise and the lack of physiological context. Respiratory effort is a key signal used in clinical scoring of OSA events, but current approaches require additional contact sensors that reduce scalability and patient comfort. This paper presents the first study to estimate respiratory effort directly from nocturnal audio, enabling physiological context to be recovered from sound alone. We propose a latent-space fusion framework that integrates the estimated effort embeddings with acoustic features for OSA detection. Using a dataset of 157 nights from 103 participants recorded in home environments, our respiratory effort estimator achieves a concordance correlation coefficient of 0.48, capturing meaningful respiratory dynamics. Fusing effort and audio improves sensitivity and AUC over audio-only baselines, especially at low apnoea–hypopnoea index thresholds.
The proposed approach requires only smartphone audio at test time, which enables sensor-free, scalable, and longitudinal OSA monitoring.
\end{abstract}

\section{Introduction}
\label{sec:intro}

Obstructive sleep apnoea (OSA) is a prevalent form of sleep-disordered breathing (SDB) characterised by repeated episodes of partial or complete upper-airway obstruction during sleep. It affects approximately 16\% of adults with higher prevalence in the older population~\cite{zasadzinska-stempniakPrevalenceObstructiveSleep2024,khokhrinaSystematicReviewAssociation2022,vaientiNarrativeReviewObstructive2024}.
Untreated OSA is associated with a range of health conditions, including cardiovascular disease, diabetes, and cognitive impairment~\cite{khayatSleepDisorderedBreathingHeart2013a}. Early detection and continuous tracking of OSA are therefore important, as timely intervention can reduce associated comorbidities and improve quality of life.

The diagnostic reference test for OSA is overnight polysomnography (PSG), which typically employs more than 12 sensors.
Sleep clinicians manually score OSA events from these multimodal signals to compute the apnoea-hypopnoea index (AHI). This metric aggregates both aponeas (complete airway collapses) and hypopnoeas (partial airway collapses) equally, defining the AHI as the total number of combined events per hour of sleep.
~~
Although accurate, PSG is costly, labour-intensive, and restricted to specialised laboratories. The use of multiple sensors can disrupt sleep, and long waiting lists further delay diagnosis. Consequently, a large proportion of adults with OSA remain undiagnosed or untreated~\cite{swanson2011sleep}. Recent evidence further highlights the limitations of one-night PSG assessments, with AHI demonstrating high night-to-night variability~\cite{malhotraMetricsSleepApnea2021,lechatHighNighttonightVariability2023}. This is clinically significant given its associations with other conditions. Longitudinal OSA monitoring provides a more reliable picture of habitual disease burden, yet current clinical pathways rarely enable frequent measurement.

Acoustic-based monitoring has been shown to be a promising non-invasive alternative. Snoring and breathing sounds carry physiological information about airway obstruction, and several studies have demonstrated feasibility of acoustic-based OSA screening. 
These studies typically address one of three tasks: SDB event detection (locating apnoea and hypopnoea events), AHI estimation (calculating the index from detected events), or severity classification (categorising subjects into healthy/mild/moderate/severe groups based on AHI cut-offs presented in Table \ref{tab:dataset_demographics}). For severity classification, Nakano et al.~\cite{nakanoTrachealSoundAnalysis2019} reported strong performance for detecting mild/moderate/severe cases (AUC 0.99), using tracheal microphones in controlled hospital environments. In contrast, \cite{romeroAcousticScreeningObstructive2022c} investigated smartphone-based acoustic monitoring at home, which achieved competitive though lower AUCs of about 0.7 for mild and 0.8 for moderate/severe OSA.
Similar declines in performance in home environments were reported in~\cite{leRealTimeDetectionSleep2023b}, who conducted a noise simulation and found that acoustic-based OSA events detection is sensitive to specific noises including air conditioning and the human voice. 
To improve the robustness of audio-based methods, many studies introduce other physiological signals or demographic data as additional information \cite{castillo-escarioEntropyAnalysisAcoustic2019, romeroObstructiveSleepApnea2023a, liCSMTCombiningSnoring2025}. 
Respiratory effort, in particular, has been shown to provide complementary information to acoustic features and significantly improve OSA screening~\cite{romeroObstructiveSleepApnea2023a}. However, current methods require additional sensors to capture effort, which limits scalability, increases cost, and reduces patient acceptability in home or longitudinal monitoring scenarios.

To overcome these barriers, we investigate whether respiratory effort cues can be inferred directly from audio. Our key hypothesis is that respiratory dynamics leave subtle imprints in snoring and breathing acoustics, and that explicitly modelling these latent cues can enhance the robustness of audio-based screening without introducing extra hardware requirements. The main contributions of this paper are: (i) We quantify the predictive relationship between snore acoustics and respiratory effort dynamics. (ii) We introduce a latent fusion strategy that integrates inferred respiratory effort with acoustic embeddings to regularise audio representations. (iii) We demonstrate that this approach improves OSA screening performance compared to audio-only baselines, while preserving the scalability and simplicity of purely acoustic methods.


\section{Data}
\label{s:data}

Experiments were conducted on a self-collected dataset~\cite{romeroAcousticScreeningObstructive2022c} acquired using a Home Sleep Apnoea Test (HSAT) device (SOMNOtouch RESP). In parallel, nocturnal audio recordings were captured in real home environments using smartphones placed near the bed. The dataset consists of 157 nights from 103 participants, whose demographic and clinical characteristics are summarised in Table~\ref{tab:dataset_demographics}. The average nightly data duration is 7 hours and the total audio recording duration is just over 1,000 hours. Compared with other PSG datasets that include concurrent audio recordings, e.g.~\cite{korompiliPSGAudioScoredPolysomnography2021a}, this dataset provides a more balanced distribution across OSA severity levels.

\begin{table}[ht]
    \centering
    \begin{tabular}{lcc}
        \hline
        \textbf{Total Participants} & \textbf{103} & \\
        \hline
        Male & 67 & 65\% \\
        Female & 36 & 35\% \\
        Age (years) & 45 ± 13 & 25 - 71 \\
        BMI (kg/m²) & 31 ± 7 & 19 - 48 \\
        \hline
        \textbf{Total Nights} & \textbf{157} & \\
        \hline
        Night duration (hours) & 7.0 ± 1.4 & 3.0 - 9.8 \\
        \hline
        Healthy: AHI $<$ 5 (nights) & 16 & 10\% \\
        Mild: 5 $\leq$ AHI $<$ 15 (nights) & 59 & 38\% \\
        Moderate: 15 $\leq$ AHI $<$ 30 (nights) & 45 & 29\% \\
        Severe: AHI $\geq$ 30 (nights) & 37 & 23\% \\
        \hline
    \end{tabular}
    \caption{Demographic and clinical characteristics of the dataset used in this study.}
    \label{tab:dataset_demographics}
\end{table}

The HSAT device records multiple physiological signals, including two respiratory effort channels derived from thoracic and abdominal movement sensors. In this study, we used the abdominal effort signal as the reference respiratory signal because it is less invasive and more comfortable for long-term monitoring.
The abdominal effort signal was sampled at 32\,Hz, while the audio recordings were sampled at 16\,kHz. To synchronise the audio with the physiological signals, the snore channel from the HSAT device (sampled at 500\,Hz), which correlates strongly with the audio signals, was used. Specifically, the audio was downsampled to 500\,Hz and aligned with the snore channel by estimating the delay via cross-correlation~\cite{romeroAcousticScreeningObstructive2022c}. After synchronisation, all model development and analysis were carried out using the original 16\,kHz audio recordings.

\section{Methods}
\label{s:methods}

All audio inputs were first segmented into overlapping 30-s windows with a 10-s shift. Each segment was transformed into 64-bin log-Mel filterbank features, with a 50-ms Hann window and a 20-ms frame shift. Each 30-s segment was therefore represented as a matrix of 1,500 frames $\times$ 64 frequency bins.

\begin{figure}[!t]
    \centering
    \includegraphics[width=.9\linewidth]{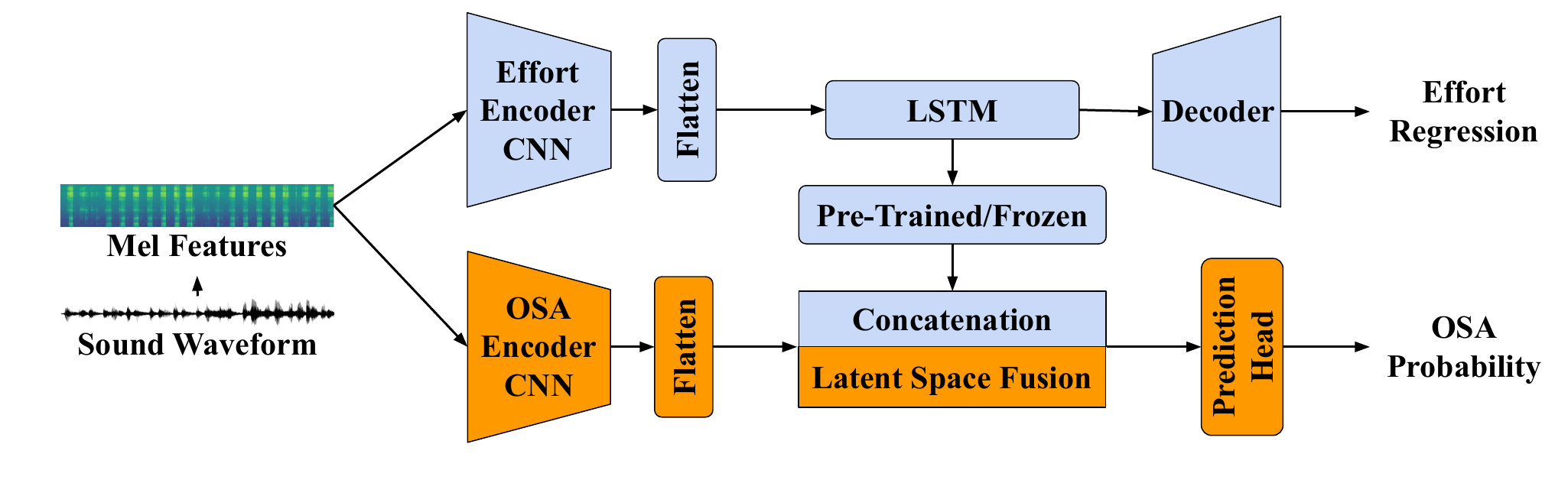}
    \caption{System diagram of the proposed latent fusion approach for acoustic-based sleep apnoea detection. A CNN-LSTM encoder extracts general acoustic embeddings and respiratory-effort-oriented latent representations from nocturnal audio. These are fused and fed into a classifier to predict apnoea/hypopnoea events.}
    \label{fig:system_diagram}
\end{figure}

Fig.~\ref{fig:system_diagram} shows the proposed two-step framework for OSA screening from audio recordings. In Step~1 (the blue part in Fig.~\ref{fig:system_diagram}), we trained an audio-to-effort estimator to infer respiratory effort directly from nocturnal audio. In Step~2 (the orange part in Fig.~\ref{fig:system_diagram}), the estimator was frozen and the final hidden state of its BiLSTM encoder was extracted as a respiratory embedding. This embedding was concatenated with the audio embedding from the audio-encoder, and the fused vector was passed through a fusion layer and a fully connected layer for classification. Importantly, no external sensors were required at test time.

\subsection{Acoustic-based respiratory effort estimation}

The proposed respiratory effort estimator consists of three main components: a convolutional neural network (CNN) feature extractor, a long short-term memory (LSTM) encoder, and a decoder with interpolation. 
The CNN feature extractor uses smaller pooling kernels than the audio-only baseline to preserve temporal resolution, converting log-Mel spectrograms from 1500 $\times$ 64 to 187 $\times$ 128 (time frames $\times$ embedding dimensions). 
The LSTM encoder then processes these 187 frames to produce hidden vectors that capture temporal respiratory dynamics. For respiratory effort prediction, a linear decoder projects each hidden vector to a single value, generating a 187-point sequence. 

The reference respiratory-effort signal is sampled at 32 Hz; each 30\,s segment therefore has 960 points. As the predicted sequence contained only 187 points, it is subsequently interpolated to match the reference effort signal length. This approach reduces LSTM training complexity while maintaining prediction of amplitude and phase consistency through interpolation. Alternative approaches, such as pooling or linear projection, do not adequately reconstruct effort dynamics.



During training, losses are computed on these upsampled predictions. A key element of training is the use of the concordance correlation coefficient (CCC)~\cite{lawrenceConcordanceCorrelationCoefficient1989} as the optimisation objective. CCC is defined as:
\begin{equation}
\label{equ:ccc}
\rho_c = \frac{2 \sigma_{xy}}{\sigma_x^2 + \sigma_y^2 + (\mu_x - \mu_y)^2},
\end{equation}
where $\mu_x$ and $\mu_y$ denote the means of the predicted and ground-truth signals, respectively, $\sigma_x^2$ and $\sigma_y^2$ are their variances, and $\sigma_{xy}$ is their covariance. CCC ranges from $-1$ (no correlation) to $1$ (perfect correlation). 

Unlike mean absolute error (MAE) and root mean square error (RMSE), CCC accounts for both correlation and bias, penalising mean and scale mismatches. We trained the model using $1 - \rho_c$ as the loss function, and report CCC as the primary evaluation metric, with MAE and RMSE provided as complementary measures.

\subsection{SDB events detection and AHI estimation}
\label{sec:AHIcalculation}

A separate audio-only CNN encoder is employed to extract audio embeddings for SDB event detection. As illustrated in Fig.~\ref{fig:system_diagram}, the proposed fusion framework integrates estimated respiratory embeddings with the audio embeddings to improve SDB event detection and AHI estimation.


Specifically, the respiratory embeddings are obtained by averaging the LSTM hidden states across time, producing a fixed-dimensional representation. This is concatenated with the audio embeddings from a baseline audio-only encoder, and the combined representation is used to predict the occurrence of apnoea and hypopnoea events. 
Finally, predicted events occurring in close succession are merged to form OSA episodes, from which night-level AHI is derived.
The AHI value is calculated as:
\begin{equation}
    \text{AHI} = \frac{\text{Number of Apnoea and Hypoponea Events}}{\text{Total Sleep Time (hours)}}.
\end{equation}

We employ a weighted binary cross-entropy loss to train the SDB event classifier, defined as:
\begin{equation}
\mathcal{L}_{\text{BCE}} = - \frac{1}{N} \sum_{i=1}^{N} w_i \left[ y_i \log(p_i) + (1 - y_i) \log(1 - p_i) \right],
\end{equation}
where $N$ is the total number of samples, $y_i$ is the ground-truth label, and $p_i$ is the predicted probability. The weight $w_i$ is defined as 
\begin{equation}
    w_i = \frac{N}{2 \times N_{c}},
\end{equation}
where $N_{c}$ is the number of samples in class $c$, used to address class imbalance between positive and negative events.

\section{EXPERIMENTAL SETUP}
\label{s:expr}

\subsection{Baselines}

We first implemented an audio-only baseline that mirrors the backbone architecture and training procedure of our proposed model, but without any respiratory-related inputs, following~\cite{romeroObstructiveSleepApnea2023a}. The network consists of three convolutional blocks with 16, 32, and 64 filters, respectively. Each block comprises convolution, batch normalisation, ReLU activation, and max-pooling. The resulting feature maps are flattened and projected to a 512-dimensional embedding, which is passed through a final fully connected layer with sigmoid activation to produce a single output score.

For reference, we also report results from an oracle system that directly uses ground-truth respiratory effort signals, as reported in \cite{romeroObstructiveSleepApnea2023a} on the same dataset. This oracle model represents an upper bound on performance achievable when high-quality respiratory measurements are available at inference time.

Finally, we evaluated our proposed latent-space fusion (LSF) model. In this setting, a frozen respiratory embedding inferred from audio is concatenated with the output of the audio encoder. The fused representation is then passed through a fusion layer followed by a multi-layer perceptron (MLP) to produce the final classification output.



\subsection{Training setup}

A 10-fold cross-validation scheme was employed for both respiratory effort estimation and AHI classification tasks. 
Data splits were performed at the subject level to ensure independence, with training, validation, and test sets following an 8:1:1 ratio.
For each fold, we first trained the respiratory effort estimation module on the training set, then froze its parameters and used the extracted latent respiratory representations fused with audio CNN features for OSA classification training. Both effort estimation and classification performance were evaluated on the held-out test set of each fold.

\subsection{Evaluation metrics}


OSA event predictions were generated using a 30-s sliding window with a 10-s shift. Adjacent predicted OSA segments were merged to form contiguous OSA events, from which the night-level AHI was computed as described in Section~\ref{sec:AHIcalculation}.

The performance of the respiratory effort signal estimator was evaluated at 30-s segment level with CCC, MAE and RMSE. Mean and standard deviation across all 10 folds are reported for each metric. We used AHI cut-offs of 5, 15, and 30 events/h to stratify OSA severity groups (healthy, mild, moderate, severe). These thresholds are widely adopted for diagnosis and treatment of OSA~\cite{kapur2017clinical}. An AHI $\geq$ 5 indicates mild OSA, and AHI $\geq$ 15 tends to lead to more arousals \cite{diederich2005sleep}. Untreated severe OSA (over 30) can lead to conditions such as hypertension and heart failure \cite{hudgel2016sleep}. 
For each cut-off, we evaluate the sensitivity, specificity and the area under the curve (AUC).

\section{RESULTS AND Discussion}
\label{s:results}

\subsection{Respiratory effort prediction from audio}

Table~\ref{tab:resp_signal_prediction} summarises the performance of respiratory effort prediction from nocturnal audio. Averaged across all cross-validation folds, the proposed estimator achieved a CCC of 0.478, with an RMSE of 1.053 and an MAE of 0.793. These results indicate that nocturnal audio contains informative cues related to underlying respiratory effort dynamics, despite the absence of direct physiological measurements.

\begin{table}[ht]
    \footnotesize
    \centering
    \begin{tabular}{l|l|l}
        \toprule
        CCC $\uparrow$ & RMSE $\downarrow$ & MAE $\downarrow$ \\
        \midrule
        0.478 $\pm$ 0.133 & 1.053 $\pm$ 0.123 & 0.793 $\pm$ 0.092 \\
        \bottomrule
    \end{tabular}
    \caption{Evaluation metrics for respiratory signal prediction: Concordance Correlation Coefficient (CCC), Root Mean Square Error (RMSE), and Mean Absolute Error (MAE).}
    \label{tab:resp_signal_prediction}
\end{table}

\begin{figure}[thb]
\footnotesize
    \centering
    \includegraphics[width=.83\linewidth]{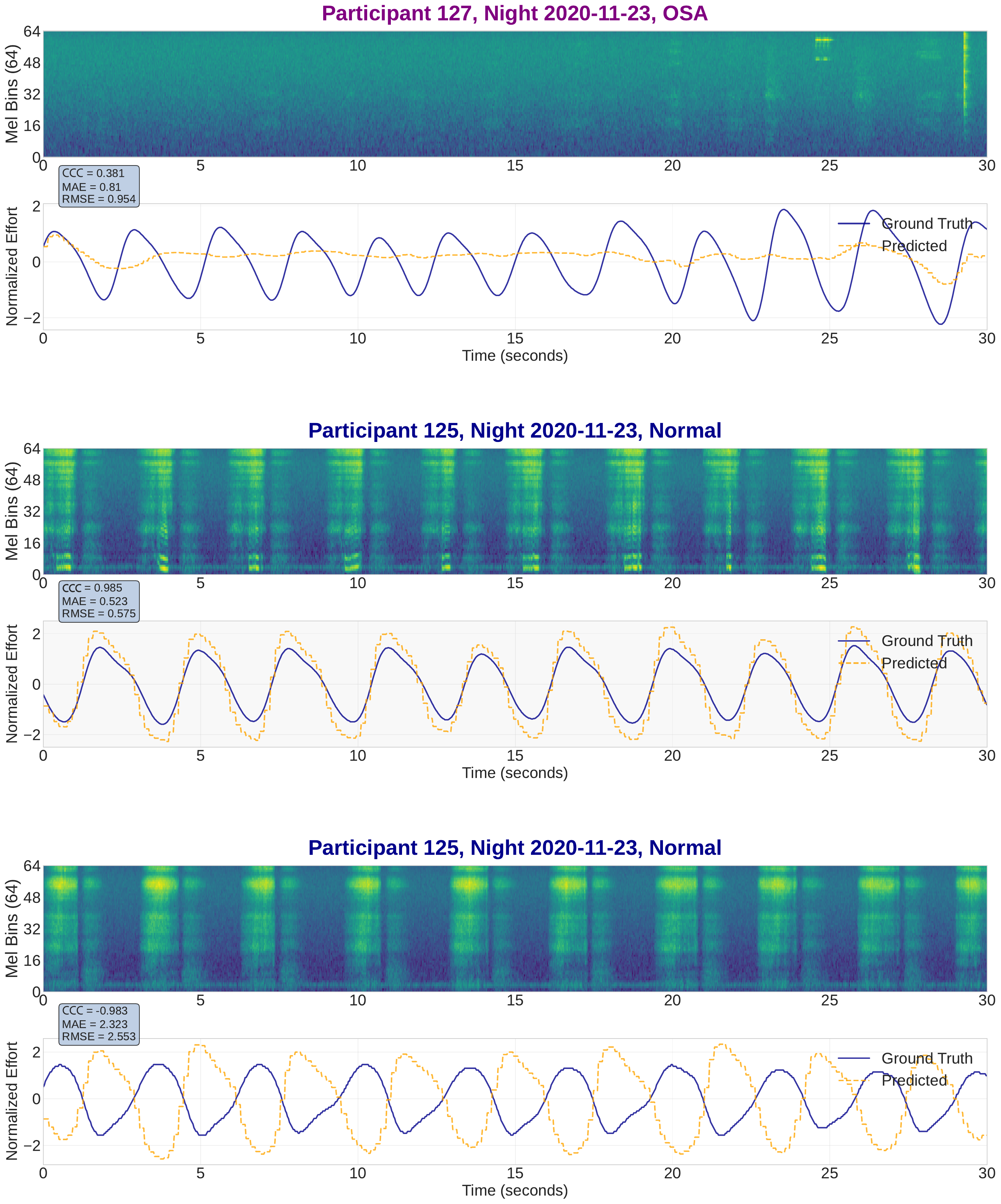}  
    \caption{Typical examples of predicted respiratory effort from audio (dashed line) compared to measured ground truth (solid line).}
    \label{fig:resp_signal_prediction}
\end{figure}

Fig.~\ref{fig:resp_signal_prediction} presents representative examples of predicted respiratory effort signals compared with the corresponding ground-truth measurements. In the first example, the estimated effort captures the overall temporal trend of respiratory events, including prolonged low-activity regions corresponding to pauses in snore-related sounds. Although the overall CCC is moderate, the predicted signal reproduces key physiological patterns. The second example shows close temporal alignment with the ground truth, whereas the third illustrates a case of phase misalignment, where effort dynamics are preserved but shifted in time. In segments with minimal audible breathing activity, the predicted effort remains largely flat, reflecting the lack of informative acoustic cues. Overall, these examples demonstrate that audio-derived estimates can serve as a meaningful, albeit imperfect, proxy for respiratory effort.

Prior work using clean speech recordings on a smaller, controlled dataset (25 hours, 26 speakers) reported higher prediction accuracy, with a CCC of 0.6 and an RMSE of 0.15~\cite{mitraPreTrainedFoundationModel2024a}. In contrast, the present results reflect the increased difficulty of modelling respiratory effort from a substantially larger and noisier sleep sound dataset collected in real-world home environments using distant smartphone microphones. The relatively higher MAE and RMSE may also be partially attributed to temporal downsampling introduced by pooling operations in the network, which reduces the temporal resolution of the predicted effort signal. Furthermore, residual temporal misalignment between audio and respiratory measurements can lead to reduced correlation in some segments, as illustrated in the lower example of Fig.~\ref{fig:resp_signal_prediction}.

\subsection{OSA severity classification}


Table~\ref{tab:ahi_classification_results} summarises OSA severity classification performance at three clinically relevant AHI cut-offs (5, 15, and 30 events/h). Results are aggregated across all cross-validation folds to reflect performance on the full dataset, as the limited number of nights per fold makes fold-wise variance estimates unstable and difficult to interpret.

At the lowest AHI threshold (5 events/h), corresponding to detection of mild OSA, the proposed LSF model with estimated respiratory effort embeddings achieves the highest sensitivity (0.88), outperforming both the audio-only baseline (0.86) and the oracle system using measured effort (0.81). This indicates that incorporating estimated effort provides complementary physiological information that is particularly valuable when OSA events are subtle and less acoustically salient. This gain in sensitivity is accompanied by a reduction in specificity compared to the oracle model (0.69 vs. 0.82), though specificity remains higher than that of the audio-only baseline (0.62).

\begin{table}[ht]
    \centering
    \footnotesize
    \begin{tabular}{lllll}
            \toprule
             & \textbf{AHI Cut-off} & \textbf{5} & \textbf{15} & \textbf{30} \\
            \cmidrule{2-5}
            & \textbf{157 nights} & \textbf{16 | 141} & \textbf{75 | 82} & \textbf{120 | 37} \\
            
            \midrule
            \multirow{3}{*}{Audio only} & Sensitivity & 0.86 & 0.81 & 0.78 \\
             & Specificity & 0.62 & 0.84 & 0.93 \\
             & AUC & 0.75 & 0.84 & 0.92 \\

            
            \midrule
            \multirow{3}{*}{\begin{tabular}[c]{@{}l@{}}Latent Fusion with\\Reference Effort \cite{romeroObstructiveSleepApnea2023a}\end{tabular}} & Sensitivity & 0.81 & 0.81 & 0.88 \\
             & Specificity & 0.82 & 0.74 & 0.87 \\
             & AUC & 0.87 & 0.86 & 0.93 \\

             \midrule
            \multirow{3}{*}{\begin{tabular}[c]{@{}l@{}}Proposed LSF Model \end{tabular}} & Sensitivity & 0.88 & 0.83 & 0.78 \\
             & Specificity & 0.69 & 0.78 & 0.93 \\
             & AUC & 0.86 & 0.88 & 0.91 \\
            
            \bottomrule
        \end{tabular}
    \caption{OSA severity classification results at various clinical AHI cut-offs (5, 15, 30 events/h). Results are aggregated across all test folds to represent the full dataset. The second row shows the class distribution between Nights $<$ Cut-off | Nights $\ge$ Cut-off.}
    \label{tab:ahi_classification_results}
\end{table}

At the moderate AHI threshold (15 events/h), the proposed LSF model again improves sensitivity over the audio-only baseline (0.83 vs. 0.81) while maintaining competitive specificity (0.78). In contrast, the audio-only baseline exhibits less stable performance, with sensitivity varying substantially across folds and dropping to as low as 0.72 in some splits. This behaviour suggests that the auxiliary effort estimation task acts as a form of regularisation, encouraging the model to learn physiologically meaningful representations rather than overfitting to dataset-specific acoustic patterns. The proposed model achieves an AUC of 0.88, the highest among all models at this mid-range threshold, where accurate discrimination between moderate and severe OSA is essential for treatment decisions.


At the highest severity threshold (AHI$\ge$30), performance differences between models diminish. The oracle system achieves the highest sensitivity (0.88), while both the audio-only and estimated-effort models attain higher specificity (0.93). AUC values converge across models (0.91--0.93), indicating that once OSA is severe, strong acoustic signatures of respiratory events reduce the added value of physiological context inferred from effort.

Overall, these results demonstrate that nocturnal audio alone contains latent physiological information sufficient to support effective at-home OSA screening. The modest but consistent gains achieved by the proposed LSF model can be attributed to several factors: (i) limited data and substantial variability in smartphone recordings and real-world room acoustics, (ii) residual temporal misalignment between audio and effort signals during zero-lag evaluation, and (iii) the use of categorical AHI thresholds, which may not fully capture the benefits of continuous respiratory representations.

Despite these challenges, the proposed method achieves AUC values close to those of systems relying on directly measured reference respiratory effort. Its key advantage lies in scalability and practicality: at inference time, only smartphone audio is required, which avoids additional sensors that may reduce patient acceptability or hinder long-term, large-scale monitoring.




\section{Conclusion}
\label{s:conc}

We presented a latent-space fusion framework that integrates respiratory effort embeddings inferred from nocturnal audio to enhance audio-only OSA screening. Our experiments demonstrate that audio-inferred effort captures meaningful respiratory dynamics and, when fused with acoustic features, provides modest but consistent gains in sensitivity and AUC at clinically relevant AHI thresholds. Compared with prior work using measured respiratory effort \cite{romeroObstructiveSleepApnea2023a}, our approach achieves smaller improvements, reflecting the challenges of noise, temporal misalignment, and reduced precision in predicted effort signals. 

A key limitation is the lack of a parameter-matched ablation analysis, which prevents a separation of the benefits of effort representations from model capacity. 
Preliminary experiments indicated that removing effort-estimation pre-training led to substantial overfitting in the proposed LSF framework at low AHI cut-offs. This suggests that the performance improvements were due to the benefits of learning an auxiliary task not increased model capacity.
However, this remains to be confirmed in future studies. 
Future work will also report additional evaluation metrics, including precision and F1-score, to better characterise the clinical utility of our approach.




\bibliographystyle{IEEEbib}
\bibliography{reference}

\end{document}